# Observing conformations of single $F_oF_1$-ATP synthases in a fast anti-Brownian electrokinetic trap

Bertram Su[a], Monika G. Düser[b], Nawid Zarrabi[b], Thomas Heitkamp[a], Ilka Starke[a], Michael Börsch*[a,c,d]

[a]Single Molecule Microscopy Group, Jena University Hospital, Friedrich Schiller University Jena, Nonnenplan 2 - 4, D-07743 Jena, Germany; [b] 3rd Institute of Physics, University of Stuttgart, Pfaffenwaldring 57, D-70550 Stuttgart, Germany; [c]Jena Center for Soft Matter (JCSM); [d]Abbe Center of Photonics (ACP), Jena, Germany

## ABSTRACT

To monitor conformational changes of individual membrane transporters in liposomes in real time, we attach two fluorophores to selected domains of a protein. Sequential distance changes between the dyes are recorded and analyzed by Förster resonance energy transfer (FRET). Using freely diffusing membrane proteins reconstituted in liposomes, observation times are limited by Brownian motion through the confocal detection volume. A. E. Cohen and W. E. Moerner have invented and built microfluidic devices to actively counteract Brownian motion of single nanoparticles in electrokinetic traps (ABELtrap). Here we present a version of an ABELtrap with a laser focus pattern generated by electro-optical beam deflectors and controlled by a programmable FPGA. This ABELtrap could hold single fluorescent nanobeads for more than 100 seconds, increasing the observation times of a single particle by more than a factor of 1000. Conformational changes of single FRET-labeled membrane enzymes $F_oF_1$-ATP synthase can be detected in the ABELtrap.

## 1. INTRODUCTION

The focus of our research for nearly 20 years has been the unravelling of conformational dynamics of the membrane-embedded enzyme $F_oF_1$-ATP synthase[1-34] which is driven by two distinct rotary motors. This ubiquitous nanomachine in the thylakoid membrane of plant cells, in the inner mitochondrial membrane of eukaryotes and in the cytoplasmic membrane of bacteria catalyzes the chemical synthesis of adenosine triphosphate (ATP) from adenosine diphosphate (ADP) and phosphate $P_i$. Therefore, it converts the chemical energy provided as a proton (or $Na^+$ in some organisms) concentration difference plus an electric potential across the membranes into mechanical energy of subunit rotation, and the rotary double motor assembly forces and synchronizes the opening and closing of the catalytic ADP and $P_i$ binding sites where ATP is synthesized[35-38].

We apply a single-molecule biophysics approach to study the bacterial $F_oF_1$-ATP synthase from *Escherichia coli* (and other membrane transporters[6, 17, 39-52]). The analysis of rotary conformational dynamics of the enzyme is achieved by single-molecule Förster resonance energy transfer (smFRET). We modified the protein by introducing cysteines or fusion of fluorescent proteins at different sites of the $F_1$ part or the membrane-embedded $F_o$ part, respectively. Thereby, we can specifically attach one fluorophore at the rotary subunits $\gamma$ or $\varepsilon$ in $F_1$, or $c$ in $F_o$. The second fluorophore for smFRET is attached to a static subunit, and rotary catalysis is observed as sequential stepwise distance changes between the two marker dyes. The FRET-labeled enzymes are reconstituted as single proteins in artificial liposomes with diameters in the range of 120 to 150 nm. Adding either Mg-ATP for ATP hydrolysis, or Mg-ADP and $P_i$ plus generating an electric potential across the membrane for ATP synthesis, respectively, results in FRET changes. Depending on the respective motor properties, *i.e.* the $F_1$ motor with the $\gamma$ / $\varepsilon$ rotor stepping in 3 steps (or, with sub-steps, 6 [53, 54] or 9 [55]), or the rotating ring of *c*-subunits in $F_o$ in 10 steps, different numbers of FRET level occur and can be discriminated by FRET efficiencies and / or dwell times of the level.

..................................................................................................................................................................

*michael.boersch@med.uni-jena.de; phone +49 36419396618; fax +49 3641933750; http://www.m-boersch.org

Confocal smFRET using freely diffusing proteoliposomes generates photon bursts which have a limited mean observation time for a proteoliposome to traverse the excitation focus and exhibit strong intensity fluctuations within a photon burst due to stochastic Brownian movement through a three-dimensional Gaussian excitation and detection profile. Both limitations, *i.e.* short time trajectories and partly low and fluctuating intensities, makes the use of likelihood estimators for FRET level or change point analysis necessary. We use Hidden Markov Models and, in part, also the *a priori* knowledge of the symmetry of the two motors to analyze the stepping behavior in a more reliable way than manual inspection and level assignment[14, 21, 22, 52]. As our attempts to immobilize the proteoliposomes with FRET-labeled $F_oF_1$-ATP synthase on a cover glass surface did not yet yield fully functional enzymes, we were looking for other options to investigate this nanomachine in solution, but on longer time scales and with constant fluorescence intensities for the sum of FRET donor and FRET acceptor dyes.

Exactly ten years ago[56-58], Adam E. Cohen and W. E. Moerner presented a novel microfluidic device that could actively compensate the Brownian motion of small particles like 20-nm polystyrene beads in water and, thereby, achieved prolonged observation times (more than 1000-fold). This device was called 'Anti-Brownian electrokinetic trap' or ABELtrap[34, 58-79]. It comprises a less than 1 μm shallow optically transparent microfluidic structure (made of PDMS, glass or quartz, respectively) to prevent diffusion in z-direction. Four access channels lead to this cross-like trapping region. In each of these access channels with a height of more than 20 μm, a Pt-electrode is positioned. Fluorescence of the molecule or nanoparticle is used to estimate its position within the trapping region in real time providing the fast feedback possibility. Depending on the actual x- and y-distances to a target position, voltages are supplied to the respective Pt-electrodes, and the fluorescent object is moved to the target position by both electrophoretic and electroosmotic forces. A. E. Cohen and W. E. Moerner quickly improved the trapping technology from EMCCD image- and software-based localization for the electro-kinetic feedback to the fastest FPGA-controlled hardware in only a few years. Nowadays, single fluorophores in solution can be trapped and their photophysics studied in unprecedented detail[73, 75, 79].

Directly after the remarkable ABELtrap oral presentation at SPIE Photonics West 2005, the author (M. B.) approached A. E. Cohen and W. E. Moerner to ask for a collaboration to study rotation of the FRET-labeled $F_oF_1$-ATP synthase at work in the ABELtrap at Stanford, and to learn how to build such an important tool for smFRET studies. Today, we report the status of our efforts to setup a fast ABELtrap in our group and to hold a single FRET-labeled $F_oF_1$-ATP synthase in an ABELtrap successfully. Therefore, this short progress report is gratefully dedicated to the ABELtrap pioneers and collaborators Adam and W. E..

## 2. EXPERIMENTAL PROCEDURES

### 2.1 Microscope setup and sample chamber preparation (PDMS microfluidics)

We built a FPGA-controlled ABELtrap as published by A. Fields and A. E. Cohen in 2011, with minor modifications[79]. Laser excitation was provided by a continuous-wave laser at 491 nm (Cobolt Calypso, 50 mW) that was attenuated to 100 to 300 μW. The laser beam was steered by a pair of electro-optical beam deflectors, EOBDs (Model 310A, Conoptics) and directed to the back aperture of the microscope objective by a lens (f = 1000 mm). We used a TIRF 100x oil immersion objective with n.a. 1.49 (Olympus) mounted in an Olympus IX71 inverted microscope, and a dichroic beam splitter (zt488rdc, AHF Tübingen) to reflect the laser and to block backscattered light from the ABELtrap. Two single photon-counting APDs recorded the fluorescence intensities in two spectral ranges from 500 to 570 nm (HQ535/70, AHF) and from 595 to 665 nm (HQ630/70, AHF), separated by an imaging beam splitter at 580 nm (BS580, AHF). Both APDs were mounted on a single 3D-adjustable mechanical stage (OWIS, Germany) that also contained the 150 μm pinhole.

Photons were recorded in parallel on two computers, *i.e.* by a FPGA card (PCIe-7852R, National Instruments) and by two synchronized TCSPC cards (SPC150 and SPC150N, respectively, Becker&Hickl, Germany), similar to our different ABELtrap setup published previously[32, 80, 81]. The FPGA Labview program from A. Fields and A. E. Cohen was slightly modified to allow for trapping on the signals of either one or on both of the two detection channels. The controls for the 3D piezo sample scanner (P-527.3CD with digital controller E-725.3CD, Physik Instrumente, Germany) were adapted in the FPGA software as well. For calibration purposes, single fluorescent nanobeads were placed on a cover glass by spin coating, and were moved stepwise by the x-y-z piezo stage.

The microfluidic PDMS design we used was published previously[82], and was similar to the original design by A. E. Cohen and W. E. Moerner. Sylgard 184 elastomer kit (Dow Corning, Farnell, Germany) was used to fabricate the PDMS chambers for the ABELtrap. Short plasma treatment of both PDMS chip and cover glass resulted in irreversible bonding.

## 3. RESULTS

### 3.1 PDMS and quartz chips, illumination pattern, calibration of the ABELtrap

The ABELtrap requires microfluidics to prevent diffusion in z-dimension. The microfluidic chips consist of either a combination of structured PDMS bonded to the cover glass, or can be made as all-quartz cells. We previously established the PDMS/glass chips using a structured silicon wafer as the template for the PDMS polymerization. We have recently started to make all-quartz cells for significantly reduced luminescence background. In Fig. 1, the design of the two masks for both the shallow trapping region (1 µm thin) and the deeper access channels with ring structure for balancing hydrostatic pressure differences (20 µm deep) are shown.

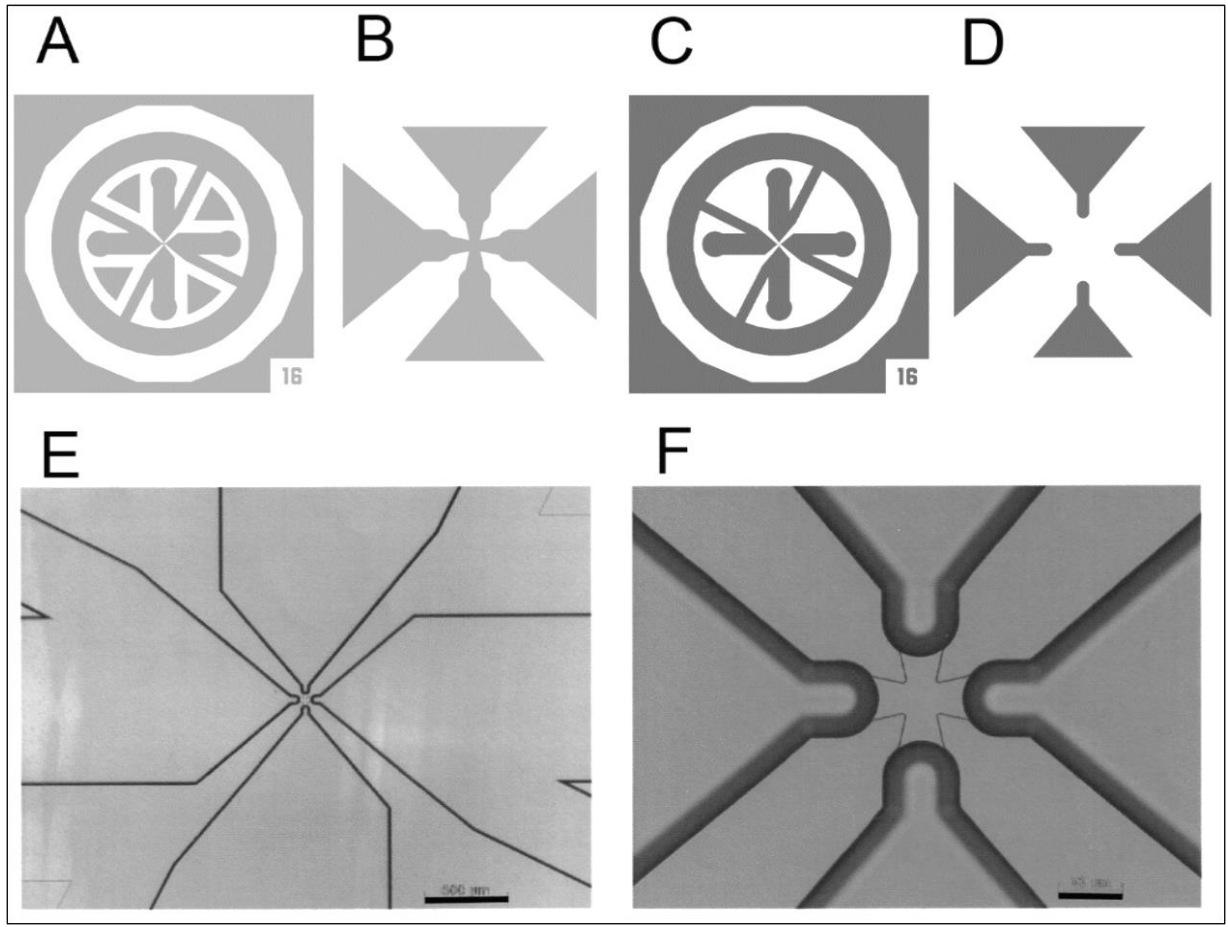

**Figure 1:** Design and realization of quartz chips for the ABELtrap by Drs. G. Mayer and T. Henkel (IPHT Jena, Germany) according to the original design by A. E. Cohen and W. E. Moerner[57]. **A-D**, design of the two masks for the thin ABELtrap region (**A, B**) and for the 20-µm-deep access channels (**C, D**) with the connecting circular structure for hydrostatic pressure balance (**A, C**, overview; **B, D**, detail of the trapping region). **E**, overview of the trapping region and the deep access channels of the etched quartz chip; black scale bar is 500 µm. **F**, detailed view of the 1-µm-shallow cross-like trapping region of the etched quartz chip; black scale bar is 50 µm.

For the ABELtrap experiments with the setup described in the following, we used our previous PDMS/glass chips. The chips have a similar design of the trapping region, but deeper access channels (~80 µm). PDMS/glass chips are cheap to produce and disposable. The drawback is a high luminescence background from both the cover glass in a spectral region above λ=600 nm, and fluorescent impurities in the PDMS.

Fluorescent nanoparticles or proteoliposomes with fluorescently labeled $F_oF_1$-ATP synthases were excited by a continuous-wave solid state laser with 491 nm emission. To scan the laser beam over an area of 2 µm times 2 µm in the sample plane that was placed within the shallow region of the ABELtrap chip, we used two electro-optic beam deflectors (Model 310A, Conoptics) with an achromatic λ/2 wave plate in between (RAC 3.2.10, B. Halle Nachf., Germany). EOBDs were driven by two fast amplifiers (Model 7602M, Krohn-Hite). Using two plano-convex lenses (f = 80 mm) we imaged the first EOBD deflection plane onto second one, which was then projected onto the back-focal plane of the 100x microscope objective (N.A. 1.49, oil immersion, Olympus) by a f = 1000 mm lens (LA1779-A, Thorlabs).

The FPGA Labview software to run the ABELtrap is distributed freely by A. Fields and A. E. Cohen[79]. We selected the circular pattern of 19 focal points and chose a diameter of the focus pattern of about 0.8 µm. Our expanded laser focus size was ~1.3 µm according to FCS measurements of diffusing dye molecules and to the EMCCD camera image of a 20-nm bead attached to the cover glass surface, respectively.

To optimize the optical alignment and to test the operation of our feedback system, we used immobilized 100-nm fluorescent beads which were scanned through the trapping region. The x- and y-feedback voltages were recorded for each position. Care was taken to ensure that the position to voltage-mapping was linear in the central part of the trapping region. We confirmed the linearity of feedback voltage *vs.* position that was sufficient in a range of roughly 0.9 x 0.9 µm (Fig. 2 A). The covered area of the laser pattern in the trapping region matched the used confocal pinhole with 150 µm diameter in the detection path.

### 3.2 Holding 20-nm fluorescent beads in aqueous solution in the ABELtrap

The performance of the ABELtrap depends on the precision of the EOBD-driven fast laser pattern (here with adjustable repetition rates for each pattern between 0 and 66 kHz), the accuracy of the estimated localization of the fluorescent particle or molecule, and the response time of the applied feedback voltages. We evaluated the ABELtrap with bonded PDMS/glass chips using diluted suspensions of fluorescent nanobeads with 20 nm in diameter, *i.e.* Fluosheres 505/515 or Fluospheres Nile Red 535/585 (with given absorbance and emission maxima, Molecular Probes) in water in the presence of 0.1% poly-vinylpyrrolidone (PVP, $M_w$ 40000, Sigma-Aldrich). Dilution and final concentration of the nanobeads were checked by FCS measurements in solution.

The mean fluorescence brightness of the Fluosheres 505/515 excited with 491 nm was very high so that we could reduce the laser power and, thereby, achieved a very low background count rate of less than 2 counts per ms (< 2 kHz). Nanobeads showed distinct brightnesses as expected, but mean photon count rates of 200 to 300 per ms were regularly detected. In Fig. 2 B, ABELtrapping of a single 20-nm bead is shown resulting in a residence time in the trap of more than 40 seconds. The time trace exhibited fluctuating intensities due to minor imperfections of the alignment of the ABELtrap.

The beginning of the intensity trace of the trapped nanobead is expanded in Fig. 2 C. At a measurement time of at 6 s, the photon count rate rose in one step from background level to the mean brightness level of this particular bead of about 120 kHz. The simultaneously recorded electrode voltage time traces fluctuated only between ± 2 V once the bead was trapped, but oscillated stochastically between ±5 V at times where no bead was trapped. Because the voltage feedback is applied for each detected photon also in the absence of a trapped fluorescent molecule (*i.e.* from background of the PDMS/glass chip, or 'dark counts' from the APDs), a corresponding apparent 'position' is estimated for each photon detection event. Because these 'positions' are immediately changing throughout the trapping area, the applied feedback voltages will cover all allowed minimum and maximum values. Symmetrical voltage fluctuations around 0 V during the residence time of the bead were interpreted as strong indications for trapping of a bead in solution and not of a surface-sticking bead, because the position of a surface-sticking nanoparticle will be likely out-of-center of the trapping region, and, as a consequence, feedback voltages will exhibit a permanent deviation from fluctuations around 0 V for for one or both x- and y-direction related electrodes.

The ABELtrap software also records the estimated deviation from the center of the ABELtrap as a time trace shown in Fig. 2 E for this nanobead. In the beginning of the time trace, *i.e.* in the absence of the bead, the apparent deviation from the center was fluctuating up to 250 nm. However, the estimated position of the trapped nanobead was deviating from the center only up to 50 nm, *i.e.* the trapped bead was confined to less than 100 nm. This confinement of a 20-nm bead was a significant improvement compared to the position fluctuations of trapped nanobeads as measured in our previous ABELtrap versions.

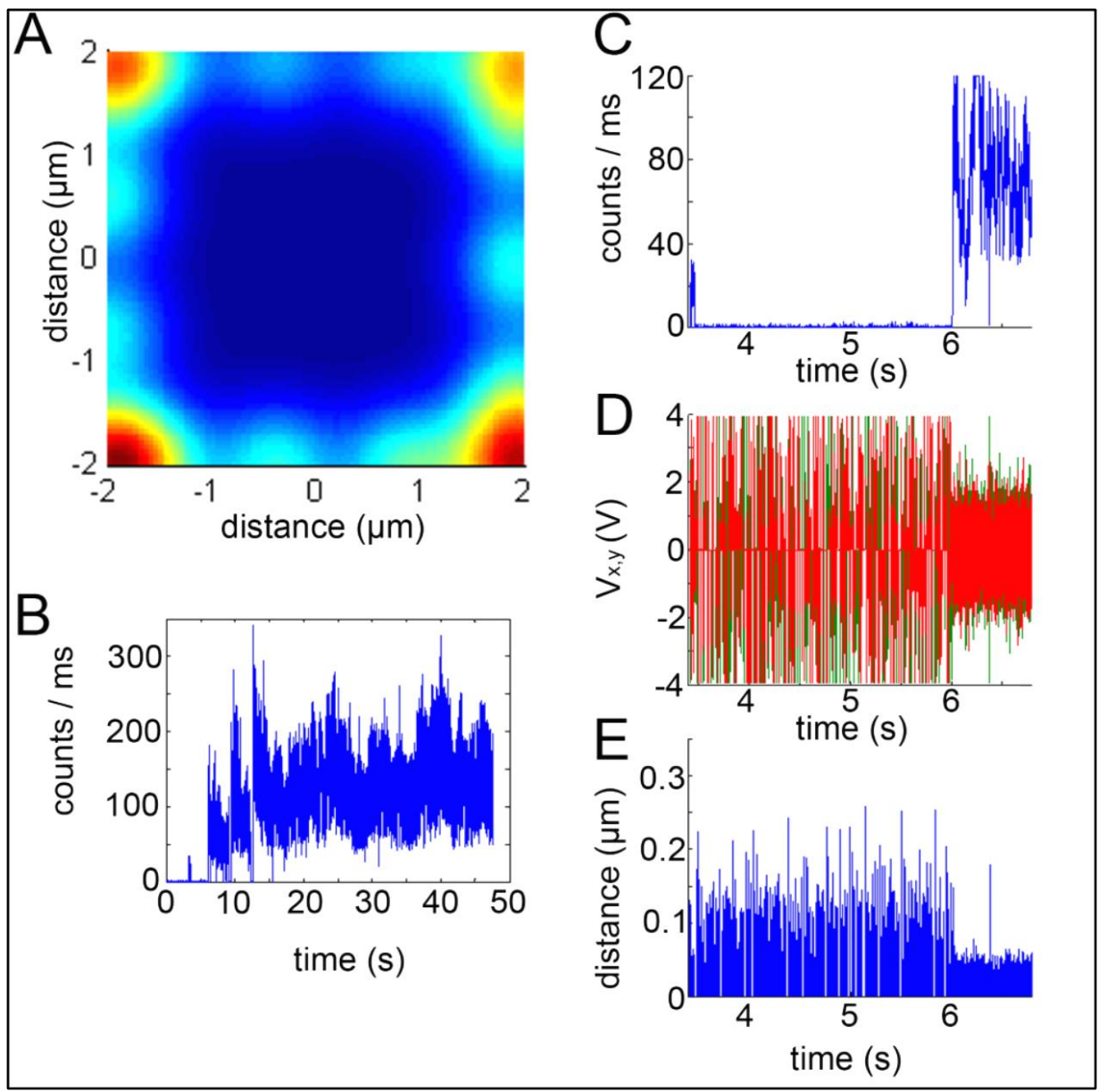

**Figure 2: A**, illumination pattern of the laser as detected by the fluorescence of a surface-attached 20-nm bead (Fluosphere 505/515). **B**, time trace of a trapped 20-nm bead in aqueous solution. The time binning was 10 ms. **C-E**, details of the time trajectories for this bead at the beginning of trapping, with 10 ms time binning. **C**, fluorescence intensity trace, **D**, electrode voltage time trace, **E**, estimated position deviation time trace plotted as distance from center, for the bead trapped at 6 s.

### 3.3 Holding proteoliposomes with labeled $F_oF_1$-ATP synthases in buffer solution in the ABELtrap

After alignment of the ABELtrap and initial performance tests with 20-nm beads, we started trapping of proteoliposomes comprising one or more Alexa488-labeled $F_oF_1$-ATP synthases. This preparation of reconstituted, labeled $F_oF_1$-ATP synthases exhibited significantly high ATP synthesis rates measured in the biochemical Luciferin / Luciferase assay, *i.e.* the cysteine mutation at the C-terminus of the membrane-embedded subunit *a* did not interfere with the catalytic activity of the enzyme nor the subsequent covalent attachment of the Alexa488 dye. In this preparation we intended to embed more than a single labeled $F_oF_1$-ATP synthase on average into the lipid bilayer of a liposome. Therefore, we expected multiple fluorophores and, correspondingly, distinct mean photon count rates for each individual proteoliposome. As seen in the fluorescence intensity time trace in Fig. 3 A, trapping a single proteoliposome resulted in different intensity levels. Some trapped proteoliposomes showed stepwise photobleaching and strong intensity fluctuations (for example, see trace at time 152.5 s). The residence time of the proteoliposomes often approached 1 second, in contrast to a mean diffusion time of only 30 ms when the ABELtrap was turned off. In some proteoliposomes the fluorescence intensities were found nearly constant (for example, see trace at time 151.5 s).

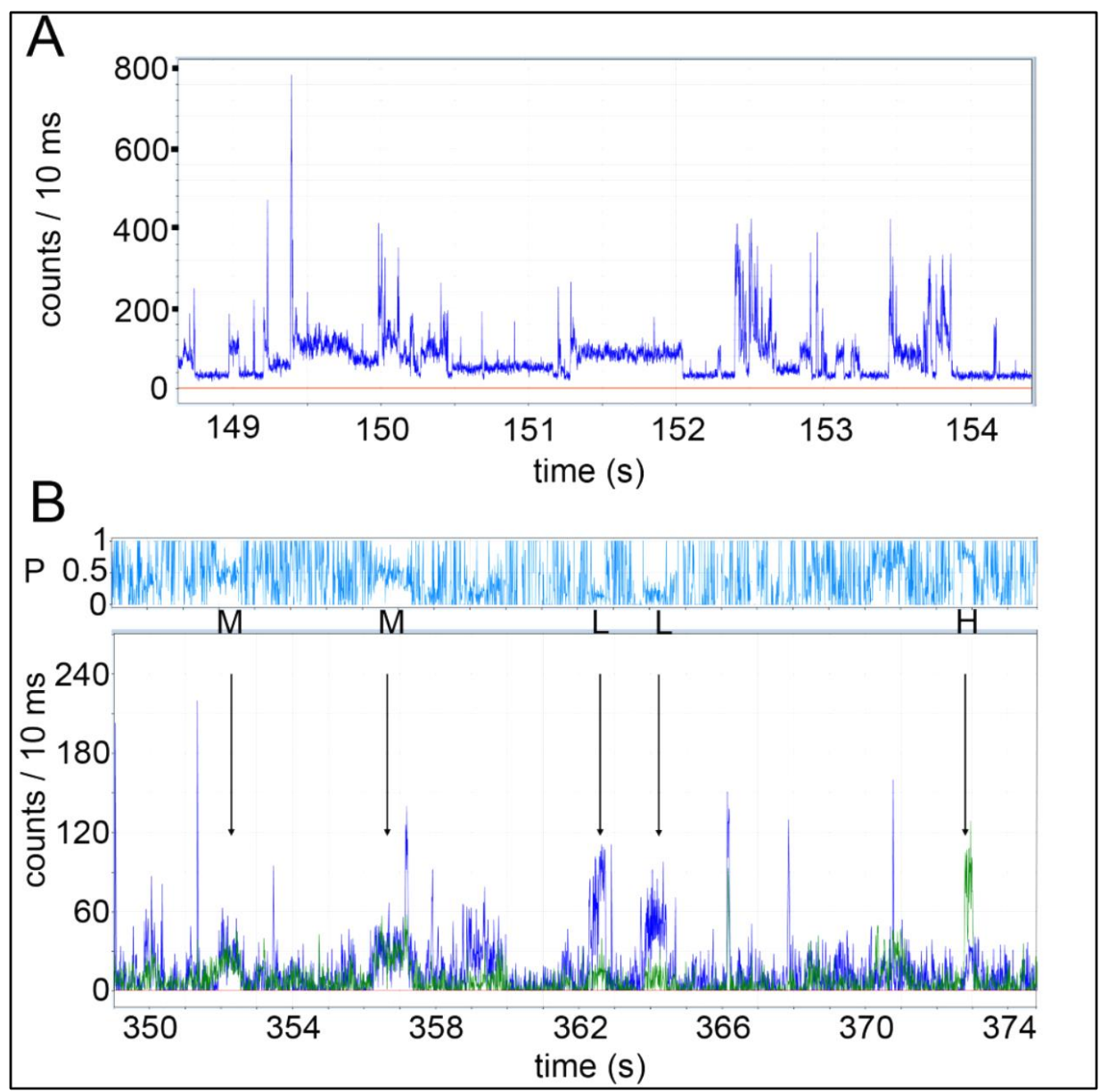

**Figure 3**: **A**, time trace of ABELtrapped proteoliposomes with one and more Alexa488-labeled $F_oF_1$-ATP synthases. Trapping was achieved using Alexa488 fluorescence photons. Note that this intensity trace is shown without background subtraction. **B**, time trace of ABELtrapped proteoliposomes with single FRET-labeled $F_oF_1$-ATP synthases with Alexa488 as FRET donor (blue trace) and Atto594 as FRET acceptor (green trace). Trapping was achieved using Atto594 fluorescence photons. Note that this intensity trace is shown with background subtraction in both channels. FRET efficiencies were calculated per time bin as the proximity factor P (time trace in upper panel) without further corrections. Mean FRET efficiencies in selected photon bursts were assigned to low FRET (L), medium FRET (M) and high FRET (H) level.

Finally we explored the possibility of trapping a single FRET-labeled $F_oF_1$-ATP synthase in a liposome with a mean diameter of 120 nm. Shown in Fig. 3 B are preliminary results of ABELtrapped Alexa488-Atto594-labeled enzymes. The time traces with donor intensity in blue ($I_D$) and acceptor intensity in green ($I_A$) were binned to 10 ms time intervals. High background count rates had to be subtracted, briefly 138 counts / 10 ms (13.8 kHz) in the donor channel and 41 counts / 10 ms (4.1 kHz) in the acceptor channel. Three types of FRET efficiencies were found for the $F_oF_1$-ATP synthases in Fig. 3 B, calculated as the proximity factor $P=I_A / (I_D + I_A)$. The low FRET efficiency (L) photon bursts originated from enzymes with an actual rotor orientation of ε that was related to a large distance between Atto594 on the ε subunit and Alexa488 on the *a* subunit. The medium FRET efficiency (M) corresponded to a different rotor orientation, i.e. yielding a shorter distance between the fluorophores. The high FRET efficiency (H) was assigned to a third ε subunit orientation within the enzyme that showed the shortest distance between the two marker dyes. Despite the high luminescence background on both detection channels, ABELtrapping of a single FRET-labeled $F_oF_1$-ATP synthase in a liposome in buffer was possible when we trapped the biomolecules on the FRET acceptor signal that showed a slightly reduced background. We added 0.1 % PVP to the buffer to prevent sticking of the proteoliposomes on the glass and PDMS surfaces of the chip.

## 4. DISCUSSION

Here we presented the current version of our ABELtrap setup in Jena based on two fast EOBDs to generate a laser focus pattern in a 2 μm x 2 μm region with up to 66 kHz repetition rate. Adapting the ABELtrap software for the FPGA by A. Fields and A. E. Cohen[79], we could achieve trapping of single FRET-labeled $F_oF_1$-ATP synthases in a liposome in buffer solution. Thereby, difficulties with surface-attachment of this nanomotor could be avoided: In future, the function of the enzyme can be studied for extended observation times in such an ABELtrap. In contrast to our previous smFRET approaches with $F_oF_1$-ATP synthases using freely diffusing proteoliposomes, the fluorescence intensities of donor and acceptor dye on a single enzyme could be analyzed quantitatively, and photophysical fluctuations could be identified helping to exclude these photon bursts from further FRET analysis.

Our previous versions of the ABELtrap, EMCCD-based[82] or AOBD-based[32, 80, 81], respectively, did not reach residence times of more than 100 seconds for fluorescent nanobeads as we have achieved now. However, the actual versions of the ABELtraps in Stanford and in Harvard have been developed even further and have implemented sophisticated learning algorithms for the diffusion and electromobility properties of a trapped molecule. This results in significantly tighter trapping and gains additional information about the hydrodynamic radius and charge (and transient changes thereof) of the trapped molecule.

PDMS/glass ABELtrap chips exhibit strong luminescence upon laser excitation with 491 nm. Therefore, quartz microfluidics are the better choice for the fast photon-by-photon feedback of the current versions of the ABELtrap. In addition, photophysical properties of both FRET donor and acceptor dyes in the respective local protein environment have to be examined carefully, and, eventually, oxygen scavenger systems have to be added to prevent early photobleaching and shortened observation times. All these methods have already been invented, optimized and applied by the ABELtrap groups of A. E. Cohen and W. E. Moerner. Thus we can expect that we will learn more und in detail about the mechanochemical mechanisms of this rotary nanomachine by observing it one after another in real time and at work in the ABELtrap.

**Acknowledgements**

We are grateful for the continuous support by Adam E. Cohen and W. E. Moerner and for providing their ABELtrap software. We want to emphasize that Thorsten Rendler, Marc Renz and Anastasiya Golovina-Leiker (Stuttgart, Germany) have built previous EMCCD-based versions of an ABELtrap and provided the design of our PDMS microfluidics, and we thank Monika Ubl (4[th] Institute of Physics, Stuttgart University, Germany) for the Si-wafers used for the PDMS microfluidics. We thank Günter Mayer and Thomas Henkel (IPHT Jena, Germany) for producing the quartz ABELtrap microfluidics; Dag von Gegerfelt (von Gegerfelt Photonics, Germany) for the loan of the Cobolt laser, Wolfgang Becker (Becker&Hickl, Germany) for the loan of TCSPC electronics, and Michael Sommerauer (AHF Analysentechnik, Germany) for the loan of high performance optical filters. This work was supported in part by the Baden-Württemberg Stiftung (by contract research project P-LS-Meth/6 in the program "Methods for Life Sciences") and DFG grants BO1891/15-1 and BO1891/16-1 (to M. B.).